# Interlayer Exchange Coupling in Semiconductor Magnetic/Nonmagnetic Superlattices

**P. Kacman, J. Blinowski [1], H. Kępa [1], T.M. Giebultowicz [1,2]**

Institute of Physics, Polish Academy of Sciences, al. Lotników 32/46, 02-668 Warszawa, Poland

[1] Physics Depart., Warsaw University, ul. Hoża 69, 00-681 Warszawa, Poland
[2] Physics Depart., Oregon State University, 301 Weniger Hall, Corvallis, OR 97311-6507, USA

The interlayer spin correlations in the magnetic/non-magnetic semiconductor superlattices are reviewed. The experimental evidences of interlayer exchange coupling in different all-semiconductor structures, based on neutronographic and magnetic studies, are presented. A tight-binding model is used to explain interaction transfer across the non-magnetic block without the assistance of carriers in ferromagnetic EuS/PbS and antiferromagnetic EuTe/PbTe systems.

## INTRODUCTION

The development of sensitive read devices consisting of magnetic layers boosted the extensive studies of electrical and magnetic properties of multilayer structures. In late 1980s, two discoveries contributed to the further increase of the potential storage capacity of magnetic materials. These were the "giant magneto-resistance", [1], and the antiferromagnetic (AFM) coupling between the ferromagnetic (FM) Fe layers separated by a nonmagnetic Cr layer, [2]. Further, it was established, [3], that it is the AFM coupling in FM layer structures that leads to the giant magneto-resistance effect (as shown schematically in Fig.1).

The ANTIFERROMAGNETIC correlations between FERROMAGNETIC layers separated by a non-magnetic spacer, which in the case of metallic structures proved to play a key role in many technological applications, such as magneto-resistive sensors and magneto-optical devices, was recently discovered also in all-semiconductor, nearly insulating, EuS/PbS superlattices (SLs), [4]. This is of great interest, since the semiconductor structures have advantages over the all-metal ones owing to the possibility of controlling the carrier concentration by temperature, light or external electric field. The coupling between FM layers was also observed in another all-semiconductor system, i.e., in multilayers made of GaMnAs, the newest generation III-V-based FM diluted magnetic semiconductor, with the (Al,Ga)As or GaAs spacers, [ 5-7 ]. While the interlayer coupling in GaMnAs-based structures, with high concentration of free carriers, can be explained, at least qualitatively, in terms of the models tailored for metallic systems (compare [ 8 ] and the references therein), the results for EuS/PbS SLs point to a different mechanism capable to transfer magnetic interactions across thick non-magnetic layers without the assistance of mobile carriers. It should be emphasized that the interlayer exchange coupling was observed in other all-semiconductor structures, i.e., in the ANTIFERROMAGNETIC short period (111) SLs: EuTe/PbTe, MnTe/CdTe, MnTe/ZnTe, [ 9-13 ], in which not only the density of carriers is several orders of magnitude lower than in metals but also another factor playing an essential role in the known theories of the interlayer coupling - the net layer magnetization - is absent.

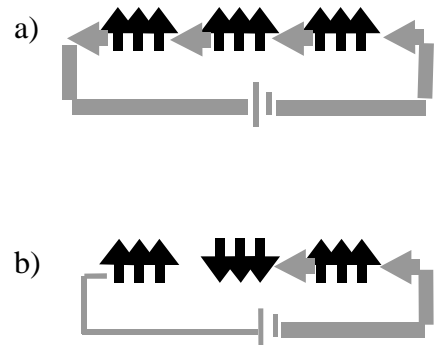

Fig.1: FM layers with the magnetic order correlated by the (a) FM and (b) AFM interlayer exchange coupling. In (b) the spin-polarized current (shaded arrows) cannot pass through the layers when the magnetic field is applied.

For the AFM layers the notions of AFM and FM interlayer coupling are not applicable. Still, in these structures there are possible two types of co-linear correlations, i.e., the identical (in-phase) and reversed (out-of-phase) spin orientations in successive layers, as shown in Fig.2.

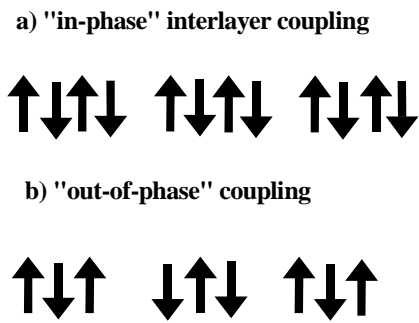

. Fig.2: The in-phase (a) and out-of-phase (b) co-linear spin structures for the correlated AFM layers

Both types of these correlations were observed in the neutron scattering experiments for SLs, in which the magnetic material was MnTe or EuTe. MnTe is a type III antiferromagnet, which in MnTe/CdTe SLs, due to the tetragonal distortion of the lattice, forms a helical spin structure. On the other hand the EuTe layers in EuTe/PbTe SLs exhibit at helium temperatures the type II AFM structure with Eu spins ferromagnetically ordered in (111) planes, which are in turn antiferromagnetically coupled one to each other.

In this review we will first present a model capable to explain the interlayer exchange coupling in both, FM and AFM, IV-VI-based semiconductor magnetic/nonmagnetic SLs. Then, we will present the neutron scattering techniques, which provide most of the convincing evidences of the interlayer exchange coupling in all-semiconductor multilayer systems. It should be noted that the only research tool capable of detecting correlations between AFM layers is the neutron diffraction. Finally we will present the results of the neutronographic studies and the comparison with the theoretical description.

## THEORETICAL MODEL

Several attempts to explain the interlayer exchange coupling in all-semiconductor structures have been reported in the literature. For the II-VI zinc-blende SLs with MnTe magnetic barriers, the exchange coupling mediated by shallow donor impurities located in the nonmagnetic quantum wells was proposed, [ 14, 15 ]. These models do not apply, however, to IV-VI structures, since in PbTe and PbS localized shallow impurity states were never detected. For the latter the interlayer spin-spin interactions mediated by valence-band electrons were suggested, [ 16]. The results obtained within the thigh-binding model have explained the origin of the interlayer correlations in the AFM EuTe/PbTe (111) SLs as well as in the FM EuS/PbS (100) SLs, with no localized impurity states. It should be noted that this model is based on the total energy calculations, which do not focus on a particular interaction mechanism, but account globally for the spin-dependent structure of valence bands.

In [ 16 ] the total electronic energies for two magnetic SLs with different spin configurations were compared: one with the magnetic period equal to the crystallographic SL period ("in-phase" interlayer coupling, i.e., identical spin configurations in successive magnetic layers - like in Fig.1(a) and Fig.2(a) ) and the other with the double magnetic period ("out-of-phase" coupling, i.e., opposite spin configurations in successive magnetic layers - Fig.1(b) or Fig.2(b) ). These calculations were performed for all studied experimentally IV-VI structures, i.e., the grown on $BaF_2$ substrate (111) EuX/PbX (where X=Te or S) SLs and the grown on KCl along the [001] crystallographic axis EuS/PbS SLs. In [ 16 ] it was assumed that the proper description of the band structure of a $(EuX)_m/(PbX)_n$ SL is reached, when the Hamiltonian reproduces in the $n=0$ and $m=0$ limits the known band structures of the bulk constituent magnetic and nonmagnetic materials, respectively. This criterion determines in principle the selection of the ionic orbitals and gives the values of nearly all parameters. It turned out that the band structures can be reproduced quite well with the Bloch functions in the form of linear combinations of s and d orbitals for Eu ions and s and p orbitals for Pb and Te (or S) ions. The nearest neighbor (NN) anion-cation interactions as well as next nearest neighbor (NNN) Te-Te (or S-S), Eu-Eu and Pb-Pb interactions had to be taken into account. Also the interactions of p-orbitals with the three NN d-orbitals belonging to the $F_2$ representation and the hybridization of anion p-orbitals with the cation f-orbitals had to be included. The values of the parameters describing all these interactions and the values of the on-site orbital energies were determined by a $\chi^2$ minimization procedure, in which the band structure was fitted to the energies known for the constituent materials in the high symmetry points of the Brillouin zone. Then, these values were used in the calculation of the difference between the total valence electron energy in the two, in-phase and out-of phase, spin configurations of the SL.

This difference can be called "correlation energy" and regarded as a measure of the strength of the valence electron mediated interlayer exchange coupling, which correlates the Eu spins across the nonmagnetic layer. The sign of the correlation energy determines the spin configuration in consecutive magnetic layers.

The main features of the obtained in [ 16 ] results are:

1) the lower energy was always obtained for the opposite orientations of the spins at the spacer borders. In the case of FM SLs this means that the interlayer coupling is AFM (compare Fig.1). For the AFM SLs this leads in the case of even $m$ to the in-phase coupling (Fig.2(a) ) whereas for odd $m$ to the out-of phase (Fig.2(b) ) interlayer coupling.

2) in all SLs, for a given number of nonmagnetic monolayers $n$ the results are essentially independent of the number of magnetic monolayers $m$, what indicates that the interlayer coupling depends primarily on the relative orientations of the spins at the two interfaces of the nonmagnetic spacer.

3) the coupling calculated in the same geometry and with the same parameters for AFM layers is approximately two times stronger than for the FM layers – the correlations do not depend solely on the spins at the interfaces;

4) in all studied SLs the correlation energy ΔE decreases monotonically, nearly exponentially, with the spacer thickness, as shown in Fig.3. The strongest and the least rapidly decreasing correlations were obtained for the FM (001) EuS/PbS SL. The comparison of the results for the two, (001) and (111), EuS/PbS SLs (in the latter case the interlayer exchange coupling was not yet observed), indicates (see Fig.3) that the valence electron mediated interlayer coupling depends strongly on the lattice geometry.

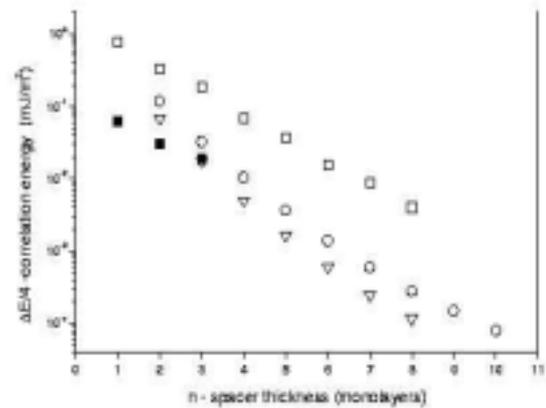

Fig.3: The absolute value of the interlayer correlation energy per unit layer surface as a function of the nonmagnetic spacer thickness for the FM (001) (squares) and (111) (triangles) EuS/PbS SLs and for the AFM EuTe/PbTe SLs (circles), as calculated in [ 16 ]. The experimental values (solid squares) for EuS/PbS (001) structures, after [ 4 ].

**NEUTRONOGRAPHIC TOOLS**

There are two powerful neutron scattering techniques that can be used for studying magnetic SLs: wide-angle diffractometry and neutron reflectometry.

In *diffraction regime*, the neutrons directly probe the correlations between individual magnetic spins in the scattering system. Therefore, this method can be used for investigating *any* type of magnetic order in a crystal - FM, AFM, or any other more complicated arrangement. If the system consists of larger "blocks" of ordered spins (e.g., of magnetically ordered layers in a SL structure), neutron diffraction is sensitive to correlations between such blocks. Let us consider a SL made up of alternating $N$ magnetic and $N$ nonmagnetic layers, each consisting of $m$ and $n$ atomic monolayers, respectively. The magnetic atoms have only two Ising-like spin states, "up" and "down". The scattering intensity $I(Q_z)$ parallel to the growth axis (z) of the SL can be obtained, from a standard equation of diffraction theory, in the form:

$$I(Q_z) \propto \left| f(Q_z) F_{s.l.}(Q_z) \sum_{l}^{N} P_l e^{ilDQ_z} \right|^2 \quad (1)$$

where $Q=(0,0,Q_z)$ is the wave-vector transfer, $f(Q_z)$ the magnetic form factor and $F_{s.l.}(Q_z)$ is the magnetic structure factor of a single layer.

The structure factor describes the peak profile that would be obtained by measuring diffraction from a single layer. It has the shape of a broad maximum accompanied by weak subsidiary maxima (the dashed line in Fig.4). For FM layers the main maxima occur at $Q_z=(2\pi/d)\zeta$ points ($\zeta=1,2,3,...$), and for AFM ones at $Q_z=(2\pi/d)\eta$ points ($\eta=1/2,3/2,5/2,...$), where $d$ is the spacing between monolayers, i.e., at the same positions, where Bragg peaks would occur for bulk crystals with the same spin structure.

In Eq.(1) by $D$ the SL period $D=(m+n)d$ is denoted. The right-side sum runs over all $N$ SL magnetic layers and the coefficient $P_l$ is +1 when the spin configuration in the $l^{th}$ layer is the same as in the $l=1$ layer, and –1 if it is reversed (compare Fig.1). The squared modulus of this sum can be divided into two terms: one describing "self-correlation" and the other the layer-layer correlations between different layers:

$$\left| \sum_{l}^{N} P_l e^{ilDQ_z} \right|^2 = N + \sum_{k \neq l} P_k P_l e^{i(l-k)DQ_z} \quad (2)$$

If there are no interlayer correlations, then the $P_l$ coefficient for successive layers takes the value of +1 or -1 in random and, for large $N$, the layer-layer correlation term disappears on statistical averaging - the spectrum has essentially the same shape as the squared single-layer structural factor.

If there are interlayer correlations in the system, the spectrum shape depends on the type of order in the layers (FM or AFM) and on how they are coupled. For ferromagnetically correlated FM layers all the coefficients have the same sign and the sum in Eq.(2) becomes equal to $\sin^2(NDQ_z/2)/\sin^2(DQ_z/2)$. This function has a sequence of sharp maxima at $Q_z = \pi\, p/D$ points ($p = 0,1,2,...$). These maxima are "enveloped" by the single layer structure factor function and the neutron diffraction spectrum has the shape shown in Fig.4(a) For the AFM interlayer coupling, on the other hand, the $P_l$ coefficients are +1 for $l$ odd and -1 for even $l$-s, and now the squared sum in Eq.(1) becomes $\cos^2(NDQ_z/2)/\cos^2(DQ_z/2)$, which has maxima at $Q_z=\pi(p+1/2)/D$ points and produces a spectrum shape as shown in Fig.4(b). Note that for FM interlayer correlations there is a central peak at the Bragg point with symmetric pairs of "satellites", whereas in the case of AFM correlations there is an *intensity minimum* at the Bragg position in between two "fringes" with equal heights. Such a clear difference in the spectrum shapes enables an easy identification of the correlation type.

For AFM layers the rules are less straightforward. Still, it can be readily checked that the spectrum shape shown in Fig.4(a) occurs in the case of the "in-phase" correlations in SLs for which *(m+n)* is an even number, and in the case of "out-of-phase" correlations for SLs with *(m+n)* odd; in other cases, the spectrum has the profile depicted in Fig.4(b).

*Neutron reflectometry.* Neutrons impinging a flat surface of a material with the refractive index n at a grazing angle $\theta$ lower than the critical angle $\theta_{crit}=[2(1-n)]^{1/2}$ are totally reflected. The reflectivity $R(\theta)$ just above $\theta_{crit}$ is a rapidly decreasing function. If on the reflecting surface a SL structure (made of two materials with different refractive indices $n_i$) is deposited, the $R(\theta)$ characteristic exhibits sharp maxima at $\theta$ values satisfying the Bragg equation $p\lambda=2D\sin\theta$. Here $\lambda$ is the neutron wavelength, *D* the SL period, and $p=1,2,3,...$ In SL made of a *magnetized* material, additional magnetic peaks occur in the reflectivity spectrum, due to the interaction between the neutron magnetic moment and the atomic momenta. This enables the determination of the type of interlayer correlations in FM SLs. For layers which are ferromagnetically coupled the magnetic and atomic structures have the same periodicity and the magnetic peaks occur at the same positions as the structural ones (Fig.5(a) ). On the other hand, the AFM coupling doubles the magnetic periodicity, and the peaks occur halfway in between the structural ones (Fig.5(b) ). It should be noted that the intensity and resolution in reflectometry is considerably better than in diffraction experiments. However, this method cannot be used for studying AFM layers with zero net moment.

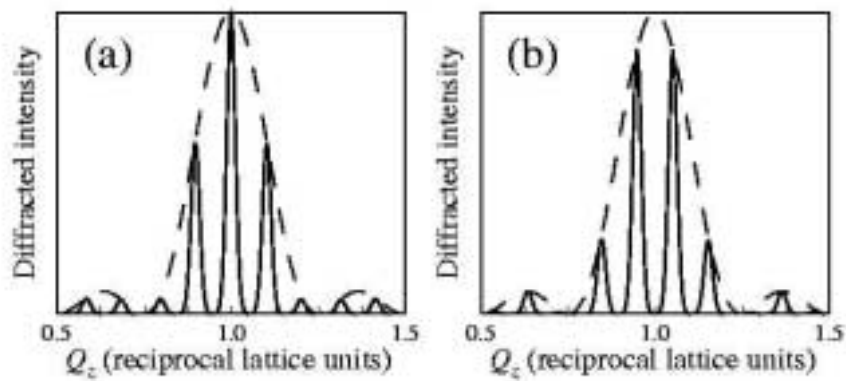

Fig.4: Diffraction profiles for a SL: (a) with FM coupling of FM layers; (b) with FM layers coupled by AFM correlations. The dashed lines represent the diffraction spectrum for an uncorrelated SL (the profile reproduces the shape of the squared single-layer structure factor).

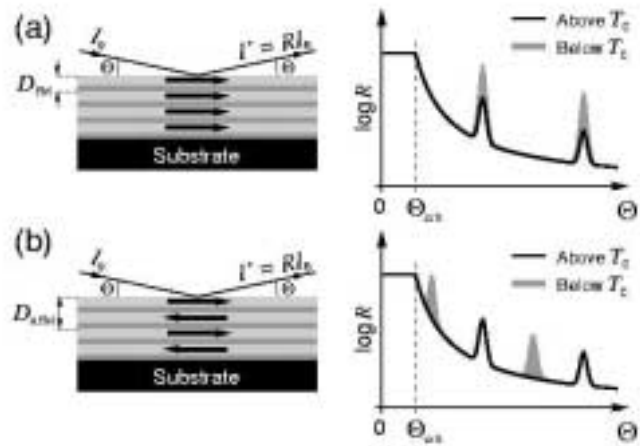

Fig.5: Reflectivity profiles from the SL structure with: (a) FM, and (b) AFM correlations between FM layers. The maxima in the solid curve are the structural Bragg peaks, the shaded profiles show the positions of the magnetic peaks arising below $T_C$ (after [ 17 ]).

## INTERLAYER COUPLING IN IV-VI-SLs

*Experimental evidences*

The neutron experiments were performed at the NIST's Neutron Scattering Center. The instruments used were BT-2 and BT-9 triple-axis spectrometers set to elastic diffraction mode, with a pyrolitic graphite (PG) monochromator and analyzer, and a 5cm PG filter in the incident beam. The wavelength used was 2.35Å and the angular collimation 40 min. of arc throughout. Additionally, a number of diffraction experiments were carried out on the NG-1 reflectometer operated at neutron wavelength equal to 4.75 Å. The latter instrument yielded a high intensity, high-resolution spectra with a negligible instrumental broadning of the SL diffraction lines.

The first neutronographic studies of the interlayer correlations in all-semiconductor structures considered the AFM SLs. Neutron diffraction measurements, carried out on a large population (~50) of $[(EuTe)_m/(PbTe)_n]_N$ SLs with many different combinations of $m$ and $n$, have revealed distinct interlayer correlation satellites in samples with $n$ up to 20 monolayers. They show that the interaction between adjacent EuTe layers can be transferred across non-magnetic PbTe spacers as thick as 70 Å. However, as can be seen in Fig.6, with increasing $n$ the satellite peaks become less sharp, while a pronounced "hump" appears underneath. The initial set of well-resolved lines gradually changes into the smooth profile characteristic for the uncorrelated structure. This indicates that the interlayer correlations weaken with the increasing thickness of the PbTe spacer. It should be noted that the strength of the coupling between these AFM EuTe layers can not be directly measured by neutron diffraction.

Recently, another structure based on the Eu chalcogenides, the ferromagnetic EuS/PbS SL grown on KCl substrates along the [001] direction, has been studied. Diffraction scans carried out at low temperatures revealed magnetic spectra with a characteristic double-peak profile (Fig.7(a) ) - a clear signature of AFM coupling between the FM layers.

This AFM interlayer coupling showed up even more clearly in reflectivity spectra (Fig.7(b) ), which exhibited sizable maxima at positions corresponding to the doubled structural periodicity of the measured specimen. Such peaks were observed for systems with the PbS non-magnetic spacer thickness up to 90Å. To confirm the magnetic origin of these peaks, reflectivity spectra were also taken with an in-plane magnetic field. Application of a sufficiently strong, external magnetic field results in full parallel alignment of the FM EuS layers; thus the AFM peak disappears, while the intensity of the peak at the structural position, corresponding to the FM spin configuration, increases.

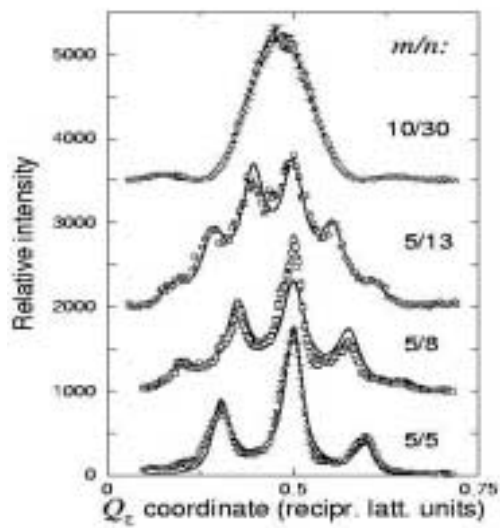

Fig.6: Diffraction peak profiles from several $(EuTe)_m/(PbTe)_n$ samples. For $n=30$ the layers of the SL are already almost completely uncorrelated. After [ 18 ].

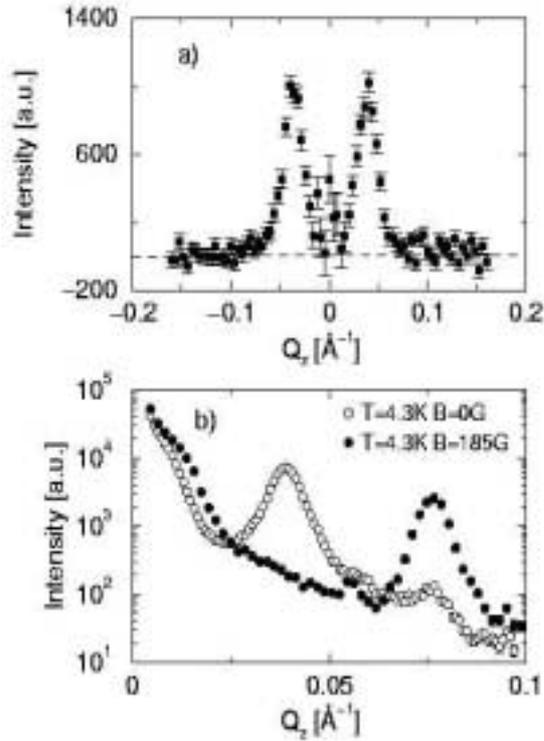

Fig.7: Diffraction (a) and reflectivity (b) spectrum from (60/23Å) EuS/PbS SL at 4.3 K. In (a) the data set taken above $T_C$ were subtracted. The double-peaked profile is characteristic for the AFM coupling. In (b) the zero field spectrum is denoted by blank points - the small structural peak corresponds to the chemical periodicity and the large one to the doubled magnetic periodicity in the SL with AFM-coupled FM layers. The external magnetic field of 185G shifts the magnetic peak to the structural position (filled points). After [ 4 ]

For the EuS/PbS SLs also the magnetic measurements, taken by a SQUID magnetometer with the in-plane field applied along the crystallographic [001] direction, were performed. The temperature and field dependences, as shown in Fig. 8, are clear indications of the presence of AFM interlayer coupling between adjacent FM layers. For SL with a given thickness of the PbS spacer, the neutron reflectivity and the magnetic measurements lead to the same value of the field needed to attain a full transition from the AFM to FM ordering of the magnetic layers (saturation field). We note that for the FM structures the saturation field provides a direct measure of the interlayer coupling strength.

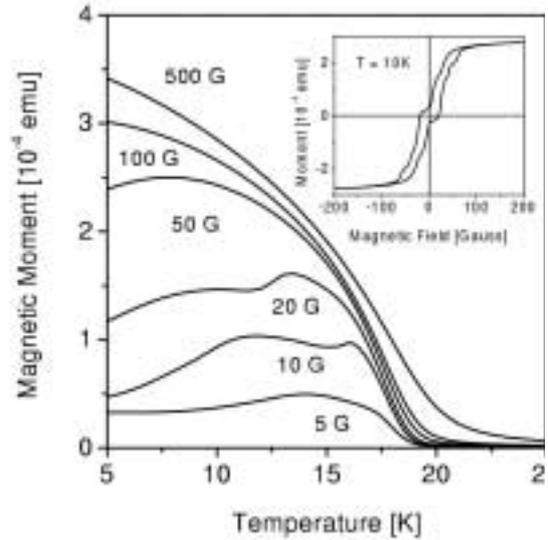

Fig.8: Temperature dependence of the magnetic moment of with the in-plane field applied along the crystallographic [100] direction. The moment versus magnetic field at 10 K is shown in the inset (after [ 4 ]).

*Comparison with the theoretical model*

For the FM (001) EuS/PbS SLs the sign of the interlayer exchange coupling and the rate of its decrease with the PbS nonmagnetic spacer thickness are in very good agreement with the predictions of the model, presented in [ 16 ], in which the interaction between the magnetic layers of the SL is mediated by valence electrons. The experimental values of the exchange constants $J_1$ estimated from the saturation fields in real structures are, however, about an order of magnitude smaller than the theoretical ones, obtained for perfect SLs (compare Fig.3). The interfacial roughness and interdiffusion, which were shown to reduce significantly the strength of the interlayer coupling in metallic structures are probably responsible for this discrepancy. The obtained theoretically features of the valence band electron mediated interlayer coupling, especially its very weak dependence on the number of spin planes in the magnetic layer, distinguish this mechanism from the AFM dipolar coupling possible in the FM multilayer structures with tiny magnetic domains. Further studies, which include preparation of samples with different thickness of the magnetic layers and different non-magnetic spacer materials, are in progress.

The comparison of the theoretical predictions for the AFM EuTe/PbTe structures with the experimental data is more complicated - in this case not only the perfect tool to measure the strength of the interlayer coupling, i.e., the saturation magnetisation, is not applicable, but also the correlated spin configurations are much more sensitive to the morphology of the SL. The information about the chosen by the coupling spin configurations in consecutive layers comes from a detailed analyzis of the positions of the satellite lines, made under an extremely strong assumption that the structures are morphologically perfect, with the same, well defined $m$ and $n$ values throughout the entire $(EuTe)_m/(PbTe)_n$ SL composed of several hundreds of periods. The observed spectra for the structures with nominally even $m$ and even $n$ reveal the preference for the in-phase spin configurations, whereas for those with odd $m$ and even $n$ they exhibit the preference for the out-of-phase configuration, both in agreement with the model predictions. None of the studied samples had even $m$ and odd $n$. For the samples with $m$ and $n$ both nominally odd, the neutron diffraction spectra seem to indicate that the in-phase configuration is preferred, contrary to the theoretical result. Still, an opposite suggestion comes from the magnetic measurements. Namely, the single period of the odd $m$/odd $n$ SLs deduced from the analysis of the neutronographic data should lead to a significant net magnetic moment of the SL - neither in EPR nor in magnetic measurements such net magnetic moment was detected, [19, 20]. Investigations of these fascinating phenomena, which include studies of EuTe/PbTe with smaller number of SL periods, i.e., with even better controlled numbers of magnetic and nonmagnetic monolayers, are in progress. Despite the current not complete understanding of the experimental data in these structures, it should be noted that the valence electron mediated interlayer exchange is up to now the only effective mechanism capable to explain the origin of the observed in EuTe/PbTe correlations between the AFM, semiconductor layers.

## ACKNOWLEDGMENTS

Support of the NATO PST.CLG 975228, FENIKS EC: G5RD-CT-2001-00535 and the NSF DMR 9972586 grants is acknowledged.